\documentclass[preprint]{aastex}

\newcommand{\iso}[2]{\hbox{${}^{#1}{\rm #2}$}}
\newcommand{\Msun}{\ensuremath{{M}_{\sun}}}

\shorttitle{Extra-mixing on the AGB}
\shortauthors{Karakas, Campbell, \& Stancliffe}

\begin{document}


\title{Is Extra Mixing Really Needed in Asymptotic Giant Branch Stars?}


\author{Amanda I. Karakas}
\affil{Research School of Astronomy \& Astrophysics, Mount Stromlo Observatory,
Weston Creek ACT 2611, Australia}
\email{akarakas@mso.anu.edu.au}

\author{Simon W. Campbell\altaffilmark{1,2}}
\affil{Departament de F\'{i}sica i Enginyeria Nuclear, EUETIB,
Universitat Polit\'{e}cnica de Catalunya, C./Comte d'Urgell 187,
E-08036 Barcelona, Spain}
\email{simon.w.campbell@upc.edu}

\and

\author{Richard J. Stancliffe}
\affil{Centre for Stellar \& Planetary Astrophysics, Monash University,
Clayton VIC 3800, Australia}
\email{Richard.Stancliffe@sci.monash.edu.au}


\altaffiltext{1}{Institut de Ci\'{e}ncies de l'Espai (CSIC), Campus
UAB, Facultat de Ci\'{e}ncies, Torre C5-parell, 2 planta, E-08193
Bellaterra (Barcelona), Spain}

\altaffiltext{2}{Centre for Stellar \& Planetary Astrophysics, Monash University,
Clayton VIC 3800, Australia}


\begin{abstract}
We demonstrate that the amount of extra mixing required 
to fit the observed low C/N and \iso{12}C/\iso{13}C 
ratios in first giant branch (FGB) stars is also sufficient to 
explain the carbon and nitrogen abundances of Galactic asymptotic/second 
giant branch (AGB) stars. We simulate the effect of extra 
mixing on the FGB by setting the composition of the envelope 
to that observed in low-mass ($M \le 2\Msun$) FGB stars, 
and then evolve the models to the tip of the AGB.  The inclusion
of FGB extra mixing compositional changes has a strong effect 
on the C and N abundance
in our AGB models, leading to compositions consistent with 
those measured in Galactic carbon-rich stars.
The composition of the models is also consistent with C and
N abundances measured in mainstream silicon carbide (SiC) 
grains. While our models cover the range of C abundances
measured in carbon stars in the LMC cluster 1846, we cannot 
simultaneously match the composition of the O and C-rich 
stars at the same time.
A second important result is that our models only match the 
oxygen isotopic composition of K and some M, MS giants,
and are not able to match the oxygen composition of 
carbon-rich AGB stars. By increasing the abundance of 
\iso{16}O in the intershell (based on observational evidence) 
it is possible to reproduce the observed trend of increasing 
\iso{16}O/\iso{18}O
and \iso{16}O/\iso{17}O ratios with evolutionary phase.
We also find that some Li production takes place during the
AGB and that Li-rich carbon stars 
($\log \epsilon({\rm Li}) \gtrsim 1$) can be produced.
These models show a correlation between increasing 
Li abundances and C.  The models cannot explain the 
composition of the most Li-enriched carbon stars, nor can 
we produce a Li-rich carbon star if we assume extra mixing 
occurs during the FGB owing to \iso{3}He destruction.
We tentatively conclude that 1) 
if extra mixing occurs during the AGB it likely only 
occurs efficiently in low metallicity objects, or when 
the stars are heavily obscured making spectroscopic 
observations difficult, and 2) that the intershell 
compositions of AGB stars needs further investigation.
\end{abstract}





\section{Introduction}

Standard stellar models\footnote{Standard in this context refers 
to models without non-convective {\em extra mixing} such as 
thermohaline mixing or rotation. This is distinguished from 
convective overshoot, which is also non-standard in stellar 
codes but is considered a natural extension of convective borders.}
predict that there is only {\em one} mixing event that 
changes the surface composition between the main sequence and
the tip of the first giant branch (FGB) -- the first dredge-up 
(FDU). The models indicate that the FDU mixes the products of 
partial hydrogen burning into the envelope, with reductions in 
the \iso{12}C/\iso{13}C ratio to $\sim 20$, and increases in 
\iso{14}N such that \iso{12}C/\iso{14}N $\sim 1.5$. The 
\iso{16}O/\iso{17}O ratio is reduced, depending on the mass of the
star, whereas \iso{16}O/\iso{18}O marginally increases
\citep[e.g.,][]{boothroyd99}. Lithium, if not destroyed during
the pre-main sequence, should be greatly reduced by the
FDU. In contrast the envelope abundance of \iso{3}He, produced 
by the pp-chains during the main sequence, is significantly
increased by the FDU.
Observations, however, do not agree with these predictions.
Low-mass giant stars have been found with \iso{12}C/\iso{13}C 
ratios of $\sim 10$ and C/N $\sim 1.0$ 
\citep{gilroy89,charbonnel94,charbonnel98}. In globular
clusters, the anticorrelation of C abundance with luminosity
is also indicative of deep mixing occurring on the FGB 
\citep[e.g., in M3;][]{gsmith02}. Furthermore, chemical 
evolution models using standard \iso{3}He yields 
\citep{dearborn96,romano03} predict present-day abundances
that are orders of magnitude greater  than the values measured
in meteorites and the local interstellar medium.  These 
observations are used as evidence that some form of chemical 
transport (known variously as extra-mixing, deep mixing, or 
cool bottom processing) is acting in low-mass red giant 
envelopes.

In recent years there has been much effort devoted to finding
the physical mechanism behind the extra mixing. Rotational
mixing \citep{charbonnel98}, gravity waves \citep{denissenkov00}, 
thermohaline mixing 
\citep{eggleton06,eggleton08,charbonnel07,stancliffe09,stancliffe10},
and magnetic fields \citep{nordhaus08,busso07b,palmerini09}
have been proposed as potential candidates. 
Parametric studies of extra mixing have also shown great promise 
in explaining the C, N, and O composition of first and 
asymptotic giant branch stars 
\citep{gsmith92,boothroyd99,nollett03}, 
although there are no compelling physical reasons for the
choice of parameters that govern the mixing and burning.

The evidence that some form of non-convective mixing 
mechanism occurs during the AGB phase is more circumstantial.
The argument put forward is similar to that for FGB 
giant stars. That is, the C and N abundances predicted
by models do not match those observed in AGB stars 
\citep[e.g.,][]{lambert86,abia97,milam09}.
In particular the \iso{12}C/\iso{13}C ratios are lower 
than predicted by standard AGB
models \citep[e.g.,][]{forestini97,karakas10}.
Also, the C and N composition of pre-solar mainstream 
silicon carbide (SiC) grains, which are assumed to form 
in the extended envelopes of carbon-rich 
(C/O $>1$) AGB stars, show a well defined distribution 
where 40 $\lesssim$ \iso{12}C/\iso{13}C $\lesssim$ 100, 
and have nitrogen isotope ratios enhanced relative to the 
terrestrial value \citep[e.g.,][]{zinner98}. In comparison,
standard AGB models predict that by the time C/O $\ge 1$
the \iso{12}C/\iso{13}C $\ge 80$, which is already close 
to the upper limit observed in AGB stars or measured
in SiC grains. A few studies have found that the only way
to match observational data of carbon-rich AGB stars is 
to artificially lower the \iso{12}C/\iso{13}C at the 
tip of the FGB to values observed in FGB stars 
\citep{kahane00,lebzelter08}. 
For example, by doing this \citet{kahane00} found the 
abundance ratios in their 2$\Msun$ model of CW Leo were
consistent with the observations. However, the authors 
did not go on to test this idea against data for a larger 
sample of Galactic carbon stars.
Lithium is also enhanced in some carbon-rich 
AGB stars. The statistics indicate that super Li-rich 
stars with Li abundances as high as 
$\log \epsilon({\rm Li}) \sim 5$\footnote{We use the
standard notation $\log \epsilon({\rm X})$ =
$\log_{10} ({\rm N_{\rm X}/N_{\rm H}}) + 12.0$, where
$N_{\rm X}$ is the number abundance of species X and 
$N_{\rm H}$ is the number abundance of hydrogen.}
are rare, accounting for only a few percent of all carbon 
stars \citep{abia93}. However a larger fraction ($\sim 10$\%) 
are considered Li-rich, with Li abundances in the range 1--2 
\citep[although some J-stars are included in this 
fraction,][]{abia93}. Standard low-mass AGB models do not
predict this variation in Li, thus extra mixing may be
needed to explain the high Li abundances in {\em some} AGB
stars \citep{abia97,uttenthaler08,uttenthaler10}. 
Intermediate-mass AGB stars can produce substantial 
quantities of Li via hot bottom burning 
\citep[HBB; e.g.,][]{boothroyd93}.  While most C-rich
stars will have a mass below the HBB limit ($M \lesssim 
4\Msun$), a few super-Li rich stars may have evolved 
from high mass AGB stars. 
These points are discussed further in \S\ref{sec:compare}.

If we assume that  thermohaline mixing (or some other 
mechanism) operates to change the envelope composition 
during the ascent of the FGB, then a natural questions 
arises: Do we need further extra mixing on the AGB 
to be able to match the observed data?  It is this question 
that we seek to address here. This question is 
particularly timely, given the recent efforts to find 
the physical mechanism driving the extra mixing on the 
FGB \citep[e.g.,][]{eggleton08}.
In this paper we argue that there is no strong need for 
extra mixing on the AGB, at least for Galactic carbon 
stars of near solar metallicity. For the moderately 
metal-poor AGB stars in the LMC and SMC the situation 
is less clear 
\citep[see also the discussion in][]{lederer09b}. 

In the current study we evolve a series of low-mass AGB models,
some of which we assume have experienced extra mixing
on the FGB. We then compare our models with observations
of C/O ratios and carbon isotopic data in a sample of Galactic carbon
stars and mainstream SiC grains, and the compositions of stars 
in the Magellanic Clusters NGC 1846 and NGC 1978. 
We include a discussion of lithium and the oxygen isotope 
ratios in AGB stars, as these species may elucidate the 
occurrence and nature of possible extra mixing in AGB stars.

\section{Stellar Models} \label{sec:models}

We evolve 1.25$\Msun$ and 1.8$\Msun$, $Z=0.01$ stellar
models from the zero age main sequence (ZAMS) to the tip of the
AGB. These masses were chosen because the 1.25$\Msun$
is likely to be close to the lower mass limit for carbon
star formation, whereas the 1.8$\Msun$ is a typical 
carbon star mass in both the Galaxy \citep{wallerstein98,abia01}
and in the LMC \citep{lebzelter07}. In regard to the
metallicities of Galactic carbon stars, \cite{lambert86}
noted that $-0.3 < [{\rm M}/{\rm H}] < 0.0$, where M is 
the total metal abundance. These
results were confirmed by \citet{abia02} who confirm 
that the metallicities of N-type carbon stars are of
nearly solar metallicity. These results suggest our 
choice of $Z = 0.01$ (or [Fe/H] $=-0.15$) is consistent
with the metallicities of Galactic carbon stars.
We calculate the structure first and perform detailed 
nucleosynthesis calculations afterward, using a 
post-processing algorithm. The details of this procedure 
and the codes used to compute the models have been 
previously described in detail \citep[e.g.,][]{karakas09}. 
In the nucleosynthesis step we use 77 species, and 
a scaled-solar initial composition where we take the 
solar abundances from \citet{asplund05} and assume
$Z_{\rm solar} = 0.015$.
Mass loss on the FGB is included using the Reimer's
formula with the parameter $\eta = 0.4$ and on the
AGB we use the \citet[][hereafter VW93]{vw93} mass-loss 
prescription. The VW93 mass-loss prescription keeps the
mass-loss rate relatively low until the beginning of the
superwind phase, at which point the mass-loss rate 
dramatically increases to a few $\times 10^{-5}\Msun$ 
year$^{-1}$. This has the effect of keeping
the envelope mass relatively high, compared to
using the Reimer's mass-loss rate on the AGB, and may
affect the depth of third dredge-up (TDU). \citet{stancliffe07}
discuss the effect of varying the mass loss
on the evolution and nucleosynthesis of a 1.5$\Msun$, 
$Z = 0.008$ model. They find that the total amount of
material that gets dredged-up to be similar for all
three mass-loss prescriptions \citep[VW93, Reimer's with
$\eta = 1$, and][with $\eta = 0.02$]{bloecker95}.
The surface C/O ratio, however, tended to 
be higher at a given time in the Reimer's model than
in the other two cases. The reason given for this is 
because models using Reimer's law typically had a 
small envelope mass at a given thermal pulse number,
It should be noted that both Reimer's and the 
Bl\"{o}cker mass loss critically depend on the choice 
of the free parameter $\eta$, the values of which 
are uncertain.

The main changes to the stellar structure code that 
have been implemented for this study are the inclusion 
of the C and N-rich low temperature opacity tables 
from \citet{lederer09},  and the inclusion of convective 
overshoot to induce the TDU in 
low-mass models. The TDU is the inward movement of 
the convective envelope into regions processed by 
partial He-burning during a thermal instability 
\citep[see][for a recent review]{herwig05}.
Some form of overshoot is required because low-mass
models computed previously with the same codes show
little or no TDU for masses $\lesssim 2\Msun$ at near
solar metallicities \citep{karakas02}. Other stellar
evolution codes also show no TDU at these masses
without some form of convective overshoot
\citep{mowlavi99a,herwig00}.   We include
overshoot following a thermal pulse by extending
the position of the base of the convective envelope by 
$N_{\rm ov}$ pressure-scale heights. To obtain a 
carbon-rich envelope we require 
$0.5 \lesssim N_{\rm ov} \lesssim 5$, 
depending on the mass of the model. 
In the  1.25$\Msun$ model we set $N_{\rm ov} = 3, 4$, 
and in the 1.8$\Msun$ model we set $N_{\rm ov} = 1, 2, 3$. 
Experimenting with the efficiency of overshoot in AGB 
models is reasonable, given our lack of understanding 
of the physics of convection in stars.  The location of 
the inner edge of the convective envelope has been shown 
to depend critically on the numerical details of a 
particular stellar evolution code \citep[see Fig.~1 in][]{frost96},
and different stellar evolution codes predict dissimilar
results for the same masses and metallicities
\citep[e.g.,][]{straniero97,mowlavi99a,karakas02,stancliffe04b}.
Efforts to match the carbon-star luminosity 
function in the Magellanic Clouds typically assume more 
efficient TDU than found in low-mass stellar 
models 
\citep[e.g.,][but see also \citealt{stancliffe05a}.]{marigo99}
Furthermore, it is still unknown if the efficiency 
of TDU is stochastic in nature, or there is one depth, 
for a given total stellar mass, H-exhausted core mass, 
and composition. 

In Table~\ref{table1} we show the initial, post-FDU, tip
of the FGB for the extra-mixing case\footnote{for the 
models without extra mixing, the abundances at the tip of
the FGB are the same as after the FDU.}, and
tip of the AGB ratios for C/O, \iso{12}C/\iso{13}C, 
\iso{12}C/\iso{14}N, \iso{14}N/\iso{15}N, 
\iso{16}O/\iso{17}O, and \iso{16}O/\iso{18}O from the
stellar models. The post-FDU values of 
\iso{12}C/\iso{13}C and C/N for the 1.25$\Msun$ model
are 27 and 1.48, respectively. The post-FDU values
for the 1.8$\Msun$ model are 22 and 1.0. These numbers
are consistent with other standard model predictions
\citep{eleid94,charbonnel94,boothroyd99}. 

To simulate the effect of extra mixing on the FGB, 
we take the envelope composition at the tip of the giant
branch and alter it such that the \iso{12}C/\iso{13}C ratio 
equals 10 \citep[from Fig.~6 of][]{charbonnel94}, 
and the C/N ratio $\approx 1$ \citep[Table~1 in][]{gilroy91}.
This is done  by decreasing the \iso{12}C abundance and
increasing the abundances of \iso{13}C and \iso{14}N,
as expected from CN cycling. The composition of the oxygen 
isotopes are not altered because the observed data for
FGB stars are consistent with model predictions 
\citep{dearborn92,boothroyd99}.
By using the lower limit for  \iso{12}C/\iso{13}C
from the observations (which show values between
10 and 18) we expect to maximize the effect on the final
AGB composition. We would expect intermediate values
to give results that fall between our extreme case and
the no mixing model. 
For the 1.8$\Msun$ case the C/N ratio is already 1.0
after FDU. After ``extra-mixing'', a reduction in \iso{12}C 
and increases in \iso{13}C and \iso{14}N result in 
C/N $\approx 0.9$ (see Table~\ref{table1}). 
This value is still within the range of observed C/N 
ratios observed in FGB stars, that show C/N ratios between
$\sim 0.8$ to 1.2 \citep{gilroy91}.

Along with simulating the effect of extra mixing on
the composition of the envelope, we perform two further
tests. The first is to increase the \iso{16}O abundance
at the ZAMS so that the initial C/O ratio $= 0.35$
(in comparison to other models which have a solar C/O = 0.5).
This is equivalent to an $\alpha$-enhanced composition, 
where [O/Fe] $\approx +0.20$. We do this because 
F and G stars in our Galaxy show moderate enhancements
of $\alpha$ elements \citep[e.g.,][]{reddy03}, and 
perhaps some of the Galactic AGB stars have similar 
moderate enhancements.

The second test is to increase the intershell 
abundance of \iso{12}C and \iso{16}O to values 
observed in the central stars of
planetary nebulae. Standard model predictions provide
values of about 25\% (by mass) of \iso{12}C in the 
intershell, and only $\lesssim 2$\% of \iso{16}O. 
Observations of post-AGB central stars suggest the 
intershell abundances are between 2--20\% for \iso{16}O 
and 15--60\% for \iso{12}C \citep{werner09}.

\section{Model predictions versus real stars} \label{sec:compare}

\begin{figure}
\begin{center}
\includegraphics[width=11cm,angle=270]{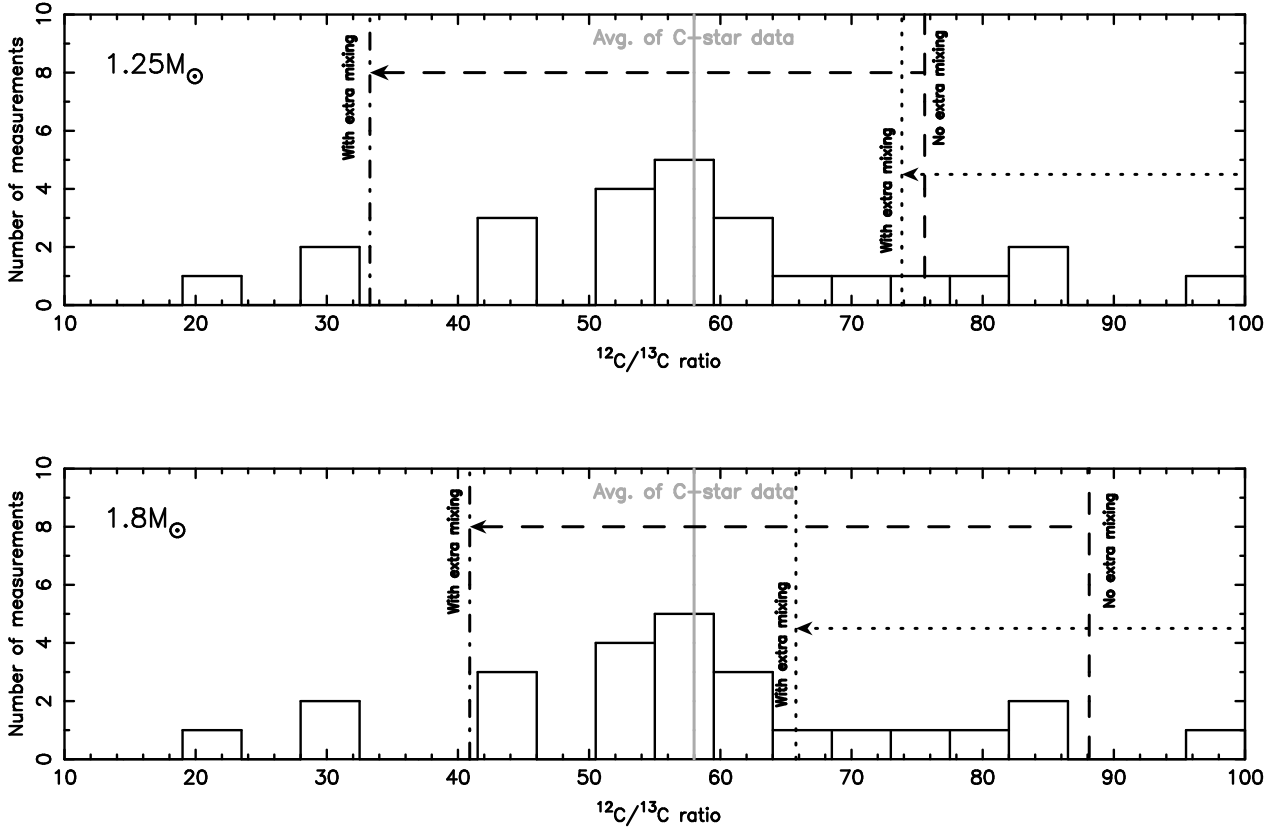} 
\caption{Histogram of observed \iso{12}C/\iso{13}C ratios 
of AGB stars from \citet{lambert86}. Data for the J-type carbon
stars have been excluded. The average value (excluding
the J-type stars) is 58, shown by the solid gray vertical line.
With the J-stars, the average is 50. The standard 
deviations are 25 and 17, with and without J-stars, 
respectively.
The top panel shows model predictions from the 1.25$\Msun$ 
model and the bottom panel for the 1.8$\Msun$ case.
In the top panel, the dashed vertical line shows the 
\iso{12}C/\iso{13}C ratio at the tip of the AGB 
for the $N_{\rm ov} =3$ model, and the dot-dashed vertical 
line shows the result when the \iso{12}C/\iso{13}C ratio 
is reduced at the tip of the FGB. The dotted line shows
the ratio for the $N_{\rm ov} =4$ model with extra mixing
on the FGB. In the bottom panel, the dashed vertical
line shows \iso{12}C/\iso{13}C ratio at the tip of the AGB 
for the $N_{\rm ov}=1$ model with no extra mixing, 
whereas the results including extra mixing are shown by
the vertical dot-dashed line. The dotted line shows the
ratio for the $N_{\rm ov}=2$ with extra mixing on the FGB.
\label{fig1}}
\end{center}
\end{figure}

In Fig.~\ref{fig1} we show the distribution of 
\iso{12}C/\iso{13}C ratios from \citet{lambert86}. 
The sample average \iso{12}C/\iso{13}C ratio for 
the C-rich stars is 58, excluding the J-stars, or 50 
with the J-stars included. The standard deviation 
on these numbers are 25 and 17, with and without the
J-stars, respectively. We choose
to use this data-set for comparison because these authors
use infra-red observations to derive the \iso{12}C/\iso{13}C 
ratios and are likely more reliable for the very red, cool 
carbon-rich  AGB stars under consideration here.
We exclude the \iso{12}C/\iso{13}C ratios for the J-type 
carbon stars, which have carbon isotope ratios close to 
the equilibrium value of $\approx 4$. The evolutionary 
origin of these \iso{13}C-rich stars has not been 
determined. Many J-stars do not show enrichments in 
$s$-process elements, in contrast to N-type stars and 
it is questionable if they are really TP-AGB stars 
\citep{wallerstein98}.  The average value of the 
\iso{12}C/\iso{13}C ratio for mainstream SiC grains 
is 57 with a standard deviation $\sigma =16$ 
\citep[using data from][]{gyngard06},  consistent 
with the average value derived for the C-rich stars
in Fig.~\ref{fig1}.

Included in  Fig.~\ref{fig1} are model predictions
from the 1.25$\Msun$ (top panel), and the 1.8$\Msun$ 
(lower panel) AGB models. 
In the top panel, the dashed vertical line shows the 
\iso{12}C/\iso{13}C ratio at the tip of the AGB 
for the $N_{\rm ov} =3$ model with no extra mixing
on the FGB, whereas the dot-dashed vertical line shows
the result when the \iso{12}C/\iso{13}C ratio is 
reduced at the tip of the FGB in line with observations.
This clearly shows the strong effect on the carbon 
isotope ratio in AGB stars when taking FGB mixing
into account. In this example the \iso{12}C/\iso{13}C 
ratio is reduced from $\sim 76$ to 33. 
The dotted vertical line 
shows the \iso{12}C/\iso{13}C ratio for the $N_{\rm ov} =4$ 
case with extra mixing on the FGB.  
This shows the effect of deeper TDU increasing the
isotopic ratio due to the extra \iso{12}C that is
dredged up.
Likewise for the bottom panel the dashed vertical
line shows the \iso{12}C/\iso{13}C ratio at the tip of the AGB 
for the $N_{\rm ov}=1$ model with no extra mixing,
whereas the results including extra mixing are shown by
the vertical dot-dashed line. Again, we can see the
significant effect of including the FGB abundance ratios.
The dotted vertical line shows the \iso{12}C/\iso{13}C 
ratio for the $N_{\rm ov} =2$ case with extra mixing
(without FGB extra mixing this model has a 
\iso{12}C/\iso{13}C ratio $> 100$, off the graph).
In summary, from Fig.~\ref{fig1} we see that the models 
considered here span the observed range of 
\iso{12}C/\iso{13}C ratios, with the exception of one
object which has a ratio of $19 \pm 6$. Two 
objects have \iso{12}C/\iso{13}C ratios
$\approx 30 \pm 5$ \citep{lambert86} but the 
uncertainties ($\pm 5$) allow these two stars to be 
spanned by the model predictions. Furthermore,
it is possible that not all C-stars will be at the 
tip of the AGB. Some of the carbon stars may have a
value between the FGB tip (after modification by
extra mixing) and the AGB tip value.  

In Fig.~\ref{fig2} we illustrate the temporal evolution 
of the C/O ratio versus the \iso{12}C/\iso{13}C ratio 
for the 1.8$\Msun$ model with $N_{\rm ov}=3$ (open symbols
and lines, refer to the caption for details). Note
that each symbol represents the model abundances after
a TDU episode. This model was chosen because it 
reaches a high final C/O ratio of 3.3 (models with 
different values of $N_{\rm ov}$ follow similar 
tracks, but finish at lower C/O ratios). It will
become apparent that once we start enhancing the O 
abundance (either at the beginning of the evolution,
or in the intershell) the final C/O ratios will be
reduced and for this comparison we require the
final C/O ratio $> 1$.  In Fig.~\ref{fig2} 
we show results from models with scaled-solar 
and $\alpha$-enhanced compositions, with and without 
extra mixing on the FGB. The labels on each line 
show the C/O and \iso{12}C/\iso{13}C ratios at the 
tip of the FGB.

Note that the models with an $\alpha$-enhanced
composition (i.e., lower initial C/O ratio) shift the 
predictions to the left, and result in a 
steeper slope and lower final C/O ratio. Also included 
in Fig.~\ref{fig2} are observational data for Galactic 
carbon stars from \citet{lambert86}, and stars in the 
Magellanic Cloud clusters NGC 1846 and NGC 1978 
using data from \citet{lebzelter08} and \citet{lederer09b}, 
respectively.
These clusters have a lower metallicity
([Fe/H] $\sim -0.5$) than considered here 
([Fe/H] $\sim -0.15$), and the AGB 
mass of NGC 1978 is smaller, at 1.5$\Msun$ \citep{lederer09b}. 
It is worth noting that the abundance tracks of the
1.25$\Msun$ model (not shown) follow almost the same
path as the 1.8$\Msun$ models, the only noticeable difference
being a final lower C/O ratio as a consequence of fewer
thermal pulses and TDU episodes. 
Even so, it is clear from Fig.~\ref{fig2} that the 
various stellar models cover most of the range of
observational data points. This agreement is especially
encouraging considering that we are using only one mass and 
one metallicity to cover the range of observed carbon-rich 
AGB stars,  where there 
is likely a spread in initial mass, composition, and 
evolutionary state.

The hypothesis put forward by \citet{lebzelter08} to explain
the composition of the NGC 1846 stars is as follows. The 
O-rich stars were matched by assuming extra mixing on the
FGB plus an $\alpha$-enhanced composition ([O/Fe] = $+0.2$), 
such that the C/O ratio $=0.2$ at the tip of the FGB, 
consistent with the observations which yield 
C/O $= 0.2 \pm 0.05$. To explain the composition of the 
two carbon-rich NGC 1846 stars near the bottom right of 
Fig.~\ref{fig2} required extra mixing on the FGB {\em and} 
AGB, but mixing only occurs on the AGB after the stars had 
become carbon rich. It is not easy to understand why 
mixing would only start once C/O $>1$ in the envelope,
given that there are several episodes of TDU before this. 
While the outer envelope (e.g., the opacities) strongly
depend on the composition, the structure of the star near 
the inner region of the envelope and the H-shell do not.
There are no noticeable changes to the structure of the 
burning shells, nor in the time between pulses as the 
star evolves from C/O $< 1$ to carbon rich.
Fig.~\ref{fig2} shows that our scaled-solar models with 
FGB extra mixing alone can explain the composition of these 
two C-rich stars.  However, the models that fit the two stars 
have a higher (0.32) C/O ratio at the tip of the FGB than 
observed in the O-rich stars by \citet{lebzelter08}. Hence 
our models cannot simultaneously fit
the composition of the O and C-rich stars at the same time.
One test of the AGB extra mixing scenario
would be the determination of the Li abundances in the
two stars.  A high Li abundance 
may imply that these stars have experienced extra mixing
on the AGB whereas a low Li abundance may rule out
the need for extra mixing.

Lastly in this section we discuss the results for
lithium from our stellar models. We will use the 
1.8$\Msun$ model with $N_{\rm ov} = 3$ as an example.
While some C and Li-rich stars with C/O ratios 
$\approx 1$ may have evolved from high-mass AGB stars 
undergoing HBB, most C-rich objects evolved from 
lower mass stars. 
The first dredge-up reduces the \iso{7}Li envelope
abundance from $\log \epsilon$(Li) from 3.14 at the
ZAMS to 1.28. It is likely that our models do not 
suffer enough Li depletion from the first dredge-up,
where observations of G--K giants suggest most stars 
have $\log \epsilon$(Li) $< 1.0$ \citep{brown89}.
During the TP-AGB the \iso{7}Li abundance in our
models increases 
as a consequence of the TDU, and reaches a maximum of 
$\log \epsilon$(Li) = 1.80. The increase of Li
mostly takes place before efficient TDU begins to
penetrate into the He-intershell (i.e., during the 
first few TPs which are weak), but after this we 
find a correlation between the Li abundance and 
carbon (and presumably, $s$-process elements).  
This could explain
the correlation found between Li and Tc abundances
by \citet{vanture07} and \citet{uttenthaler10}. 
We note that the final Li
abundance of this model would be considered Li-rich
according to the definition of \citet{abia93}
i.e., $\log \epsilon$(Li) $\ge 1.0$. 
If our model had suffered extra depletion on
the FGB, then it is possible  that the Li abundance
would not have reached  $\log \epsilon$(Li) $\ge 1.0$
during the TP-AGB.
We test this by reducing the envelope \iso{7}Li
abundance at the tip of the FGB to 
$\log \epsilon$(Li) = 0.0 \citep[close to the 
center of the Li abundance distribution for 
G--K giants, Fig.~3 from][]{abia93}. We also
reduce the envelope \iso{3}He abundance by a 
factor of 10, because this isotope would also be 
destroyed by whatever process destroys the lithium.
While we find a similar trend as before, with Li 
increasing prior to the start of deep TDU 
and then slowly thereafter, the final value is
0.76, less than the minimum for Li-rich carbon stars.

\subsection{The oxygen isotope ratios} \label{sec:oxygen}

In this section we compare our models to oxygen
isotopic measurements of FGB and AGB stars. In particular,
we are interested to see if standard AGB models can match the
observed composition, and if not, do we need to resort to
extra mixing on the AGB? We focus on oxygen
isotopic data from the stellar spectra of evolved stars and do 
not compare to the oxygen composition of presolar oxide grains. 
We do this because the oxygen isotope data for stars spans 
a smaller range than measured in grains, with few stars found in the 
region occupied by Group 2 grains, and no stars found in the
region occupied by Groups 3 and 4 \citep[Fig.~9 from][]{nittler97}.
If these grains formed in the outflows of AGB 
stars then there should be stellar counterparts. Or perhaps 
the grains formed in circumstellar envelopes around 
AGB stars much more evolved (and heavily obscured) than the MS, S
and C-type giants observed by Harris and collaborators. 
If so, it is curious that the C isotope data measured in SiC
grains span the same range as measured in AGB stars while
the O isotopic data in oxide grains do not.

In Fig.~\ref{fig3} we show the \iso{16}O/\iso{18}O ratio
versus the \iso{16}O/\iso{17}O ratio from a sample of 
G and K giants, barium stars, and carbon-rich AGB stars.
References are provided in the figure caption. Error bars 
are not included but uncertainties are large; we refer to 
\citet{harris87} for a detailed discussion.
Included in the figure are our predictions from the
1.25$\Msun$ and 1.8$\Msun$ scaled-solar models. 
We note that we are using models without extra mixing
on the FGB, but the results would not change because 
we assume that the oxygen isotope ratios are not altered
by this mixing.
We also indicate in the figure the solar oxygen isotope
ratios, which provide the starting position for our
models (the dotted lines). 
In Table~\ref{table1} and Fig.~\ref{fig3}, it can be seen 
that the FDU reduces the \iso{16}O/\iso{17}O 
ratio to $\sim 330$ in the 1.8$\Msun$ model, and 1800 in the 
1.25$\Msun$ case, as a consequence of CNO cycling that 
synthesizes some \iso{17}O. In contrast, the \iso{16}O/\iso{18}O 
ratio increases slightly as a consequence of \iso{18}O 
destruction. These numbers are reasonably consistent with
the O isotope ratios measured in FGB stars \citep{harris84}.

It is also evident in Table~\ref{table1} and 
Fig.~\ref{fig3} that the predicted oxygen isotope ratios 
at the tip of the AGB do not match 
the isotopes measured in C-rich AGB stars. In particular, the 
oxygen isotope ratios derived from observations of AGB 
stars are notably larger than the model predictions.
Is this caused by AGB extra mixing? This is unlikely, as the 
changes caused by extra mixing would shift the 
\iso{16}O/\iso{17}O and \iso{16}O/\iso{18}O ratios in a 
similar direction as the FDU. That is, the 
abundance of \iso{17}O would be enhanced, \iso{18}O 
destroyed, and \iso{16}O would remain largely unchanged.
This would result in increases in \iso{16}O/\iso{18}O and
decreases in \iso{16}O/\iso{17}O. Instead, both ratios 
appear to increase in the stars with evolution. 
\citet{herwig00} also noted that the oxygen isotope ratios 
measured in AGB stars are correlated with each other, and 
seem to be correlated with the observed carbon and $s$-process 
element abundances. Hence it is likely that the increase 
of the oxygen isotope ratios with evolutionary
state can be attributed to the TDU.

We now explore another possibility to explain the 
discrepancy. The operation of the TDU mixes material 
from the He-burning
shell, that has been exposed to partial He-burning, to the surface.
The composition of the intershell is mostly \iso{4}He and
\iso{12}C (with typical compositions 75\% and 25\%, respectively).
Partial He-burning also synthesizes small quantities of \iso{16}O 
and \iso{22}Ne, at the level of a few percent each at most.
Could it be that the amount of \iso{16}O in the intershell 
is higher than predicted by standard models?

Support for this idea comes from both observational and
theoretical fronts. The composition of hot post-AGB stars show
C abundances between 15--60\%, O abundances between 2--20\%, and
Ne abundances of about 2\% \citep{demarco01,wesson08,werner09}.
The theoretical models of \citet{herwig00} with exponentially
decaying overshoot at all convective borders find O abundances
in the intershell of $\sim 20$\%. 
The overshoot penetrates into the top of the C-O core,
increasing the C and O content of the intershell during a 
convective thermal pulse. Furthermore, key He-burning
reactions rates (e.g., \iso{12}C($\alpha,\gamma$)\iso{16}O)
are uncertain and could allow for higher \iso{16}O 
intershell abundances.

The next question is how do these intershell compositions 
effect the envelope composition? To test this idea, we 
synthetically compute the
envelope abundances of the 1.25$\Msun$ model with $N_{\rm ov}=4$, 
and 1.8$\Msun$ model with $N_{\rm ov}= 3$. All tests assume
a scaled-solar initial composition and that extra mixing has 
taken place on the FGB.
From the detailed stellar structure computations, we have the 
total mass, H-exhausted core mass, and mass dredged
into the envelope as a function of thermal pulse number.
Assuming a given set of intershell abundances we then calculate the
envelope abundance as a function of pulse number. 
In Table~\ref{table2} we show the \iso{12}C and \iso{16}O
intershell abundances used (in mass fractions), and the resultant
surface abundances at the tip of the AGB for the C/O, 
\iso{12}C/\iso{13}C, \iso{16}O/\iso{17}O, and 
\iso{16}O/\iso{18}O ratios. The surface nitrogen isotope 
ratio is assumed not to change as a consequence of increased
oxygen in the intershell. These results are illustrated
in Fig.~\ref{fig3} for the oxygen isotope ratios, and
in Fig.~\ref{fig4} for the C/O and \iso{12}C/\iso{13}C ratios.
In Fig.~\ref{fig3} the open squares 
show the result of using an intershell composition of 20\% 
for O in the 1.8$\Msun$ model. Note that the oxygen
isotope ratios do not significantly shift from the 
post-FDU values for the 1.8$\Msun$ model assuming a standard 
intershell composition.
The open circles show the evolution of the O isotope ratios
in the 1.25$\Msun$ case with a standard intershell composition
(26\% C and 0.08\% O), whereas the open triangles show the
result of using 20\% O. The two models with O-rich intershells
show a similar trend of increasing oxygen isotope ratios
with evolution as seen in the observations.

The comparison is not perfect, however, but the tests performed here
are simple and rely on many assumptions. First, we have assumed
constant intershell abundances, whereas detailed models show that
the He-intershell abundances of \iso{12}C and \iso{16}O depend
on the total mass, core mass, initial composition, and time during
the convective thermal pulse. Second, we have only enhanced the
abundance of \iso{12}C and \iso{16}O, and ignored changes to the
minor \iso{17}O and \iso{18}O isotopes. This is a valid assumption
given that the C-O core below the He shell contains only trace 
quantities of these isotopes. Also, we have only considered 
models of two masses, whereas real carbon-rich AGB stars in the 
Galaxy likely evolve from initial masses between about 1.2$\Msun$ 
to $\sim 3\Msun$. Certainly models of stars of mass between 
1.25$\Msun$ and 1.8$\Msun$ may help fill in the gap between 
the model predictions in Fig.~\ref{fig3}. We also do not vary 
the rates of the uncertain \iso{17}O($p,\gamma$)\iso{18}F and
\iso{17}O($p,\alpha$)\iso{14}N reactions, which play an important
role in setting the minimum \iso{16}O/\iso{17}O value 
reached after the FDU. The new evaluations of
the \iso{17}O $+ p$ rates by \citet{chafa07} have reduced
uncertainties compared to previous estimates, but these still
range between $\sim 20$\% to a factor of two, depending
on temperature.  

In Fig.~\ref{fig4} we show the results of an increasing
C and O intershell composition on the evolution of the C/O 
and \iso{12}C/\iso{13}C ratios from the 1.8$\Msun$ model with
$N_{\rm ov}= 3$. The solid line shows model predictions assuming 
a standard intershell composition (24\% C and 0.6\% O) and extra
mixing on the FGB. The dashed and dot-dashed lines show the
result of increasing the \iso{16}O content of the intershell
to 10\% (dashed) and 15\% (dot-dashed), which has the effect 
of shifting the final C/O ratio to lower values.  We note 
 that the extra \iso{16}O from the intershell
helps to keep the final C/O ratio within the observed range.
The dotted line illustrates results that use the composition of 
PG 1159-035, which has 48\% C and 17\% O \citep{jahn07,wesson08},
in the intershell.
That the model predictions in Fig.~\ref{fig4} fit
the data for Galactic carbon stars remarkably well 
may indicate that current AGB models predict too little
\iso{16}O in the intershell. This uncertainty warrants
further investigation.

\section{Discussion and concluding remarks} \label{sec:discussion}

We now return to the original question posed in the
Introduction: Do we need extra mixing on the AGB? 
If we simply seek to explain the C/O and \iso{12}C/\iso{13}C
ratios in Galactic AGB stars then we have found the 
answer to be {\em no}.
The inclusion of extra mixing on the FGB followed 
by a standard thermally-pulsing AGB evolution 
can produce C and N compositions consistent with values
observed in AGB stars and measured in pre-solar 
SiC grains. 

But what about Li in carbon-rich Galactic AGB stars?
The existence of stars that are both C- and Li-rich 
has led some authors to conclude that extra mixing 
occurs in some fraction of AGB stars 
\citep[e.g.,][]{abia97}. The frequency of 
carbon stars rich in Li is  $\lesssim 10$\%, 
implying that the majority of carbon stars are 
Li normal or Li poor ($\log \epsilon$(Li) $< 1$).
Furthermore, there is evidence that the Li 
production is
tied to thermal pulses (and possibly the TDU) in  
AGB stars, as indicated by the correlation between
AGB stars rich in Li and enriched in the unstable
$s$-process element Tc \citep{vanture07,uttenthaler10}.
We found that 
lithium is produced during the TP-AGB phase of the 
1.8$\Msun$ model with deep TDU, and we find a 
correlation between increasing Li and carbon in
the envelope (and presumably also $s$-process 
elements).  That our models can reproduce the 
correlation between He-shell products and Li found 
by \citet{uttenthaler10} is compelling, but 
further study into Li nucleosynthesis is required.
Our model can only become a Li-rich carbon star
when we assume that no extra mixing has taken
place during the FGB. This model then has a 
higher \iso{12}C/\iso{13}C and C/O ratio than
observed. If we assume extra mixing has taken
place on the FGB, the star still produces Li 
but is no longer considered ``Li-rich'', hence 
extra mixing would be required during the AGB to
provide the Li. However we restate that
only $\sim 10$\% of carbon stars are observed 
to be Li-rich, which indicates that efficient 
extra mixing during the AGB is {\em not} 
required in the majority of these objects. 
We note that this is consistent with our findings
for C and N. The spread in Li observed in 
carbon-rich AGB stars may be explained by the
fact we have used one stellar mass
and metallicity for our tests whereas Galactic
carbon stars will have a range of stellar masses,
some of which are predicted not to experience
any extra mixing during the FGB \citep[i.e.,
$M \ge 2\Msun$][]{eggleton08}. Also, it  is possible
that stars experience some variation in the
mixing efficiency, with the Li abundance able to
increase or decrease depending on the details 
of the mixing and burning \citep{wasserburg95}.

\begin{figure}
\begin{center}
\includegraphics[width=11cm,angle=270]{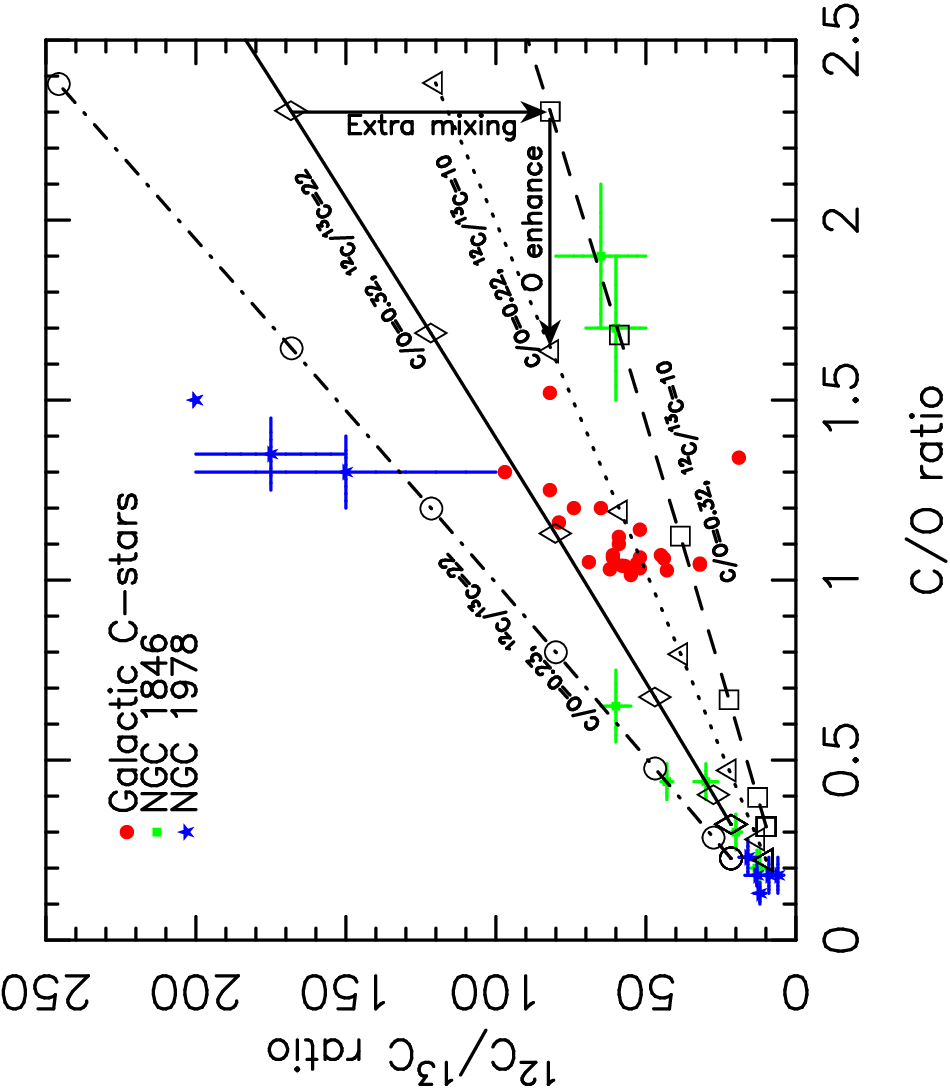} 
\caption{The C/O ratio versus \iso{12}C/\iso{13}C.
Observational data of Galactic C-stars from 
\citet{lambert86} are shown as solid circles, 
data for NGC 1846 from \citet{lebzelter08} as 
solid squares, and data for NGC 1978 from \citet{lederer09b} 
as solid stars. Error bars are included for the 
two Magellanic Cloud clusters, except for one star which
has a lower limit of 200 for \iso{12}C/\iso{13}C.
Open symbols (connected by various line types) 
indicate model predictions from the 1.8$\Msun$ model with 
$N_{\rm ov} = 3$ after each TDU episode.  
Predictions are shown for two models with a scaled 
solar composition, with and without extra mixing, and 
for two $\alpha$-enhanced models, with and without 
extra mixing. Labels indicate the C/O and \iso{12}C/\iso{13}C 
ratios at the tip of the FGB, with the standard model
indicated by the solid line.
\label{fig2}}
\end{center}
\end{figure}

\begin{figure}
\begin{center}
\includegraphics[width=11cm,angle=270]{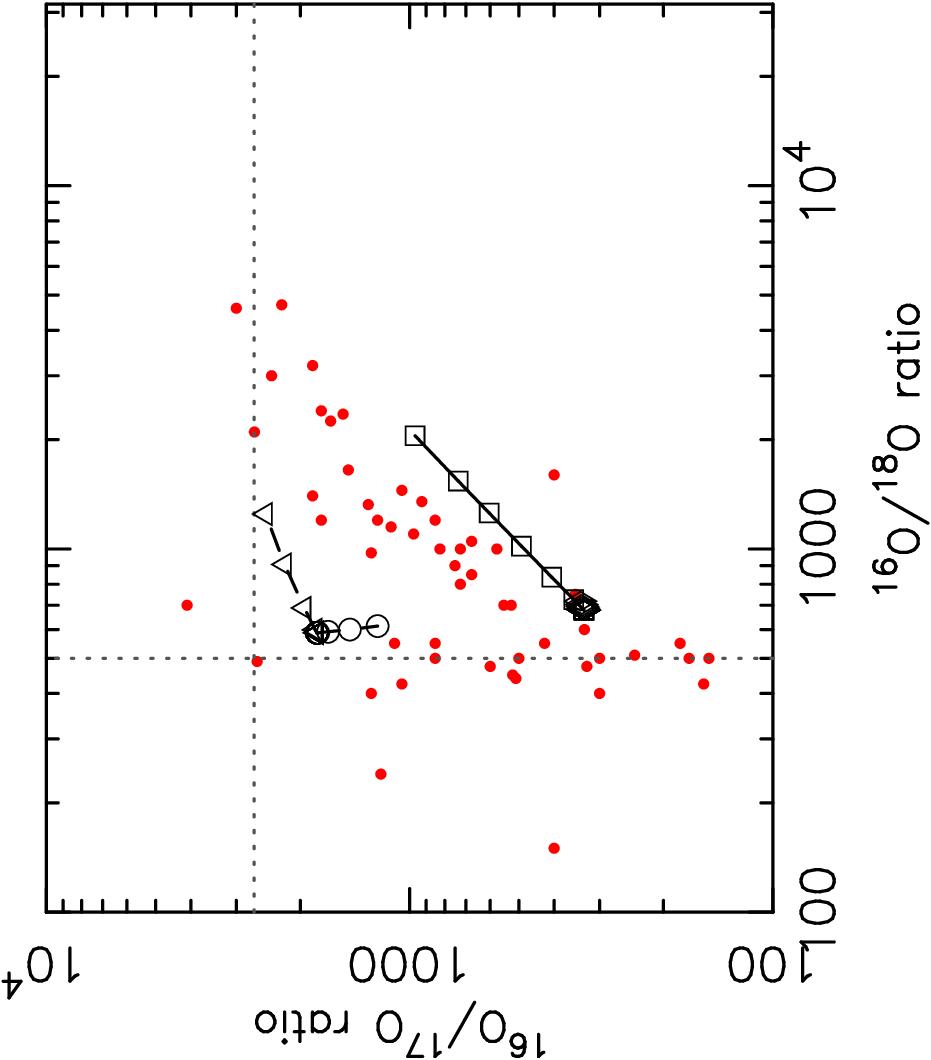} 
\caption{The \iso{16}O/\iso{18}O ratio versus \iso{16}O/\iso{17}O.
The solid points show observational data from \citet{harris84}, 
\citet{harris85a}, 
\citet{harris85b}, and \citet{harris87} for G and K giants, 
barium stars, MS and S-type stars, and carbon stars. Error bars are 
not shown but uncertainties are large; see e.g., \citet{harris87} 
for details. The dotted lines show the solar values for the
oxygen isotope ratios, which are the initial values used in
the stellar models. The open diamonds show the 
surface composition after each TDU from the 1.8$\Msun$ model 
with $N_{\rm ov}=3$ assuming a standard intershell composition,
and open squares an \iso{16}O intershell abundance of 20\% (by mass).
The open circles show results from the 1.25$\Msun$ model with 
$N_{\rm ov} = 4$ assuming a standard intershell composition, 
and open triangles assuming an \iso{16}O intershell abundance 
of 20\%. 
\label{fig3}}
\end{center}
\end{figure}

\begin{figure}
\begin{center}
\includegraphics[width=11cm,angle=270]{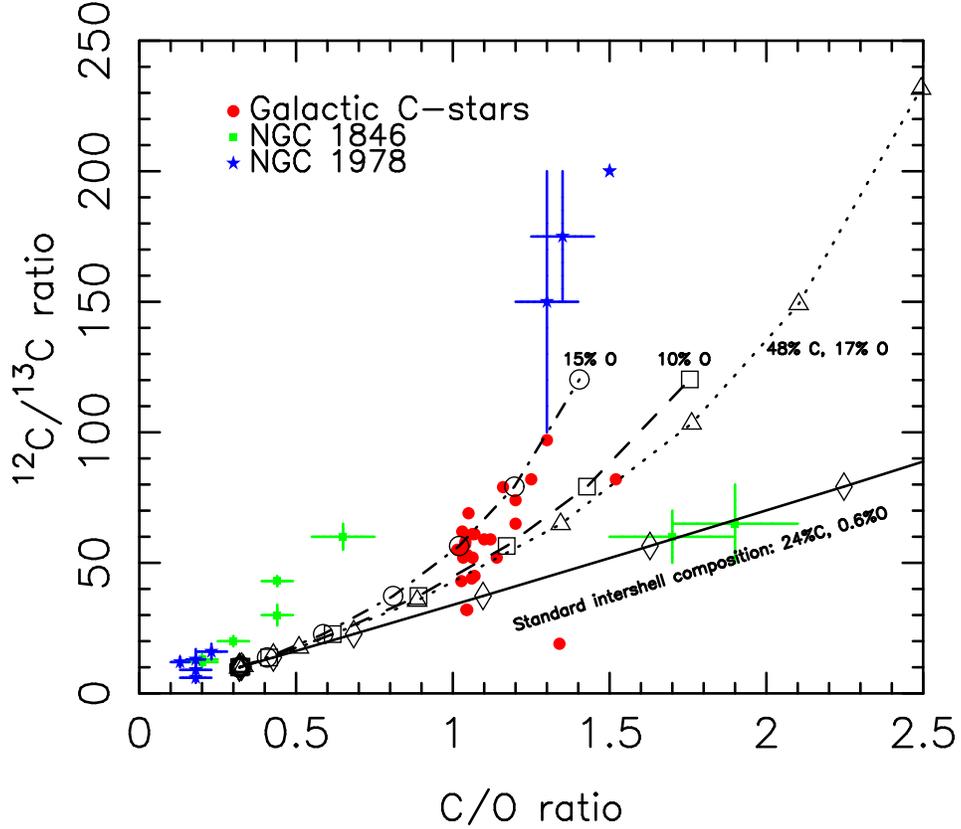} 
\caption{The C/O ratio versus \iso{12}C/\iso{13}C.
Filled symbols are the same as in Fig.~\ref{fig2}.
Open symbols indicate model predictions from the 
1.8$\Msun$ model with $N_{\rm ov} = 3$ with extra mixing on the FGB.
In this plot we show the result of varying the intershell 
\iso{12}C and \iso{16}O abundances. The dotted line shows the
result of using the composition of PG1159-035 which has
48\% C and 17\% O. The dashed and dot-dashed lines shows
the results of increasing the \iso{16}O abundance from 
the standard $\approx 0.6$\% in the intershell to 10\% (dashed)
and 15\% (dot-dashed), while keeping \iso{12}C constant 
at 24\%.  
\label{fig4}}
\end{center}
\end{figure}

The low C/O ratios present in C-rich AGB stars as 
compared to theoretical models 
(see e.g., Fig.~\ref{fig2}) have been previously 
explained by allowing for extra mixing on the AGB or 
by the use of C-rich low-temperature opacities. Extra 
mixing would reduce the C abundance but keep the 
elemental O abundance approximately
constant \citep[e.g.,][]{nollett03}.
\citet{marigo02} showed that the use of C-rich 
low-temperature opacities leads to an increase of 
the mass-loss rate and therefore limits the final 
C/O of the model star. In our detailed models we 
use the latest low-temperature C-rich opacities 
from \citet{lederer09} and we still find higher 
final C/O ratios than spanned by the observations, 
although these models depend on the assumed 
efficiency of convective overshoot. 
Even so, our models suggest that there
are only two ways to reduce the C/O ratio in
AGB models:  Variations in the intershell abundances 
or extra mixing on the AGB. 
In Fig.~\ref{fig4} we show that models with high
O intershell abundances are able to produce C/O ratios
within the range measured in Galactic carbon stars.
With a reduced TDU, these same models may also be able
to account for AGB stars rich in $s$-process elements 
but with spectral types M and MS, which have C/O $< 1$. 
We have shown extra mixing need not be the only 
explanation for these observations. This was also 
discussed by \citet{herwig00} although his models were 
more massive (3,4$\Msun$) than considered here and 
likely more massive than most carbon stars.

The case for extra mixing on the AGB is much
stronger at low metallicity. The LMC clusters NGC 
1846 and NGC 1978 both have a metallicity lower 
than solar, although NGC 1846 is slightly more 
metal poor with [Fe/H] $= -0.49$ 
\citep{grocholski06}, whereas NGC 1978 has 
[Fe/H] $= -0.37$ \citep{mucciarelli07}. 
To explain the composition of the O and C-rich 
stars in NGC 1846, extra mixing was proposed to 
occur on the FGB and AGB, but only on the AGB once 
the stars had become carbon rich. In contrast, 
extra mixing on the AGB does not seem to be required 
in the LMC cluster NGC 1978, where the C-rich stars have
high \iso{12}C/\iso{13}C ratios ($\ge 150$) 
\citep{lederer09b}. Why the evolution of the C-rich
stars in the cluster NGC 1978 should be so different
to those in the cluster NGC 1846 is puzzling, given
the small difference in metallicity.
From the trend of the models in 
Fig.~\ref{fig4}, we speculate that an intershell
enhanced in both \iso{12}C and \iso{16}O 
may be able to explain the abundances of the LMC
cluster NGC 1978. Stellar models of the appropriate
mass and metallicity as the AGB stars in that cluster
are needed to explore this possibility in more 
detail.

At lower metallicities ([Fe/H] $\lesssim -2$) 
very low \iso{12}C/\iso{13}C ratios are measured in
carbon-enhanced metal-poor 
(CEMP) stars \cite[e.g.,][]{beers05,sneden08}, as 
are high Li-enrichments \citep{aoki08,thompson08}.
The origin of the CEMP stars with enrichments of 
$s$-process elements has been suggested to be mass 
transfer from a previous AGB companion. This is 
supported by observations that suggest that 
CEMP stars with $s$-process elements are all members 
of binary systems \citep{lucatello05}. 
Extra mixing on the FGB of the donor AGB star 
would not be enough to explain the low carbon isotopic 
compositions measured in the CEMP stars 
\citep{sivarani06,campbell08}, especially those 
that are still on the main sequence or sub-giant branch 
and whose \iso{12}C/\iso{13}C ratios 
cannot therefore have been reduced by extra mixing 
during their own FGB.
\citet{stancliffe10} has suggested that 
thermohaline mixing can explain the Li enrichment, 
but the models do not have low enough 
\iso{12}C/\iso{13}C ratios.

The oxygen isotope ratios measured in evolved
giants also do not support a case for extra mixing on
the AGB. The observed increase in the 
oxygen isotopic composition with evolutionary state can
be  explained by the action of the TDU and higher O 
intershell abundances than predicted by standard models
\citep[see also the discussion in][]{herwig00}.
Increasing O is motivated by both theoretical models
of AGB stars as well as observations of hot post-AGB 
objects. One issue with this comparison are the large 
uncertainties in the measurements of oxygen isotope
ratios in evolved stars. Updated measurements with smaller
uncertainties would certainly help constrain this scenario.
The oxygen isotope data for pre-solar oxide grains
has been used as supporting evidence for 
the existence of efficient
extra mixing on the FGB and AGB \citep{zinner05,nollett03,nittler97,
wasserburg95}. However, the most extreme grains 
fall outside of the region determined for stellar 
spectra. If these grains have an AGB origin, then 
either they are formed in stars too dusty and heavily
obscured for spectroscopic analysis, or the stars
that form such grains are rare and not included in
observational samples. One test  that would help to confirm 
(or rule out) such mixing would be to derive accurate
measurements of oxygen isotope 
ratios for large samples of O-rich and C-rich AGB stars
covering a range of metallicities. 
This is desperately needed if we are to sort out the
conflicting results for oxygen, especially given the
consistency found for carbon between pre-solar SiC grains
and AGB stars. 

The models and tests we performed have many uncertainties, 
including the depth and efficiency of the third dredge-up, 
which may be stochastic in nature, as well as 
mass loss, opacities, and reaction rates. We have 
circumvented the uncertainties related to the origin 
of the extra mixing in FGB stars by adopting the
observed abundance ratios at the tip of the FGB for
our AGB models. We have also simply assumed that all 
stars experience extra mixing on the FGB such that the
value of the \iso{12}C/\iso{13}C ratio after mixing 
is equal to 10, whereas the observations
by \citet[e.g.,][]{gilroy91} show a variation between
$\sim 10$ to 25. Our results indicate that the 
observational data  for C/O and 
\iso{12}C/\iso{13}C could in fact be fit better by 
assuming values between these extremes. 
Furthermore, the question of AGB intershell
abundances may be affected  by some of these uncertainties. 
Is convective overshoot the only way to obtain large O 
intershell abundances? Instead, could variations
in key reaction rates (e.g., 
\iso{12}C($\alpha,\gamma$)\iso{16}O) allow us to 
synthesize more \iso{16}O in the intershell?
Clearly further investigation into the intershell 
abundances of AGB stars is warranted.

\acknowledgments

We thank the referee for many useful and instructive 
comments that have helped to improve this paper.
AIK thanks Orsola De Marco for discussions 
about post-AGB stars, and acknowledges support from the 
Australian Research Council's Discovery Projects funding 
scheme (project number DP0664105), and is grateful for the 
support of the NCI National Facility at the ANU.  
This research was partially supported by a grant 
from the American Astronomical Society.
SWC acknowledges the support of the Consejo Superior de
Investigaciones Cient\'{i}ficas (CSIC, Spain) JAE-DOC
postdoctoral grant and the MICINN grant AYA2007-66256. 
RJS acknowledges support from the Australian Research 
Council's Discovery Projects funding scheme 
(project number DP0879472).


%% in parentheses, after the keyword, but they will not be verified.

%{\it Facilities:} \facility{Nickel}, \facility{HST (STIS)}, \facility{CXO (ASIS)}.



\clearpage

%% the file, uncomment the \plotone, \plottwo, and
%% If you need a layout that cannot be achieved with \plotone or
%% \plottwo, you can invoke the graphicx package directly with the
%% \includegraphics command or use \plotfiddle. For more information,
%% please see the tutorial on "Using Electronic Art with AASTeX" in the
%% documentation section at the AASTeX Web site,
%% http://www.journals.uchicago.edu/AAS/AASTeX.
%%
%% The examples below also include sample markup for submission of
%% supplemental electronic materials. As always, be sure to check
%% the instructions to authors for the journal you are submitting to
%% for specific submissions guidelines as they vary from

%% This example uses \plotone to include an EPS file scaled to
%% 80% of its natural size with \epsscale. Its caption
%% has been written to indicate that additional figure parts will be
%% available in the electronic journal.

%% Here we use \plottwo to present two versions of the same figure,
%% one in black and white for print the other in RGB color

\clearpage

\clearpage

\clearpage

\clearpage

%% If you are not including electonic art with your submission, you may
%% mark up your captions using the \figcaption command. See the
%% No more than seven \figcaption commands are allowed per page,
%% so if you have more than seven captions, insert a \clearpage

%% Tables should be submitted one per page, so put a \clearpage before

\clearpage

\begin{table}[t]
\begin{center}
\caption{Initial composition, post-FDU, tip-FGB for the extra-mixing
case, and tip of the AGB abundance results for C, N, and O. 
\label{table1}}
\vspace{1mm}
\begin{tabular}{crrrrrrrr}
\tableline\tableline
Position & $N_{\rm ov}$\tablenotemark{a} & Mix?\tablenotemark{b} &
C/O\tablenotemark{c} & \iso{12}C/\iso{13}C &
\iso{12}C/\iso{14}N & \iso{14}N/\iso{15}N & \iso{16}O/\iso{17}O &
 \iso{16}O/\iso{18}O \\ \tableline
\multicolumn{9}{c}{1.25$\Msun$, $Z = 0.01$ scaled-solar model.} \\ \tableline
Initial & -- &  -- & 0.500  & 89.0 & 3.60 & 272.0 & 2680 & 500 \\
post-FDU & -- & --   & 0.387  & 26.6 & 1.48 & 798.1 & 1801 & 587 \\ 
tip-FGB  & -- &  Yes & 0.335  & 10.0 & 1.00 & 960.2 & 1801 & 587 \\
tip-AGB & 3.0  & No  & 1.077 & 75.6 & 4.01 & 884.0 & 1240 & 610 \\
tip-AGB & 3.0  & Yes & 1.015 & 33.3 & 3.13 & 1005  & 1229 & 607 \\
tip-AGB & 4.0  & No  & 2.213 & 160  & 8.38 & 876.9 & 1225 & 615 \\
tip-AGB & 4.0  & Yes & 2.165 & 73.8 & 6.87 & 1039  & 1223 & 615 \\
\tableline
\multicolumn{9}{c}{1.8$\Msun$, $Z = 0.01$ scaled-solar model.} \\ \tableline
Initial & -- & --   & 0.500  & 89.0 & 3.60 & 272.0 & 2680 & 500 \\
post-FDU & -- &  -- & 0.326  & 22.1 & 0.99 & 1269  & 331.0 & 678 \\
tip-FGB  & -- &  Yes & 0.320 & 10.0 & 0.91 & 1295  & 331.0 & 678 \\
tip-AGB & 1.0  & No  & 1.244 & 88.1 & 3.87 & 1309  & 325.8 & 694 \\
tip-AGB & 1.0  & Yes & 1.252 & 40.9 & 3.66 & 1261  & 328.2 & 698 \\
tip-AGB & 2.0  & No  & 1.960 & 142  & 6.20 & 1292  & 331.0 & 703 \\
tip-AGB & 2.0  & Yes & 1.966 & 65.8 & 6.31 & 1265  & 330.4 & 703 \\
tip-AGB & 3.0  & No  & 3.318 & 246  & 10.7 & 1217  & 338.9 & 717 \\
tip-AGB & 3.0  & Yes & 3.318 & 120  & 10.4 & 1253  & 338.2 & 718 \\
\tableline 
\multicolumn{9}{c}{1.8$\Msun$, $Z = 0.01$ $\alpha$-enhanced model.} \\ \tableline
Initial & -- & --   & 0.350 & 89.0 & 3.60 & 272.0 & 3803 & 707 \\
post-FDU & --  & -- & 0.230 & 22.1 & 0.99 & 1269  & 344.0 & 962 \\
tip-FGB  & --  & Yes & 0.226  & 10.0 & 0.91 & 1293  & 344.0 & 962 \\
tip-AGB & 3.0  & No  & 2.378 & 246 & 10.7 & 1207 & 345.3 & 1000 \\
tip-AGB & 3.0  & Yes & 2.380 & 120 & 10.4 & 1238 & 345.3 & 1000  \\
\tableline \tableline

\tablenotetext{a}{This is the overshoot parameter that governs the
efficiency of the TDU.}
\tablenotetext{b}{This indicates if extra-mixing is assumed to occur
on the first giant branch.}
\tablenotetext{c}{All abundance ratios are by number.}

\end{tabular}
\end{center}
\end{table}

\begin{table}[t]
\begin{center}
\caption{AGB abundance results for C and O for 
models with various intershell compositions. The 1.25$\Msun$
model has $N_{\rm ov} = 4$ while the 1.8$\Msun$ model has
 $N_{\rm ov} =3$.
\label{table2}}
\vspace{1mm}
\begin{tabular}{llrrrr}
\tableline\tableline
\iso{12}C$_{\rm is}$$^{a}$ & \iso{16}O$_{\rm is}$$^{a}$
&  C/O$^{b}$ & \iso{12}C/\iso{13}C & \iso{16}O/\iso{17}O & \iso{16}O/\iso{18}O \\ \tableline
\multicolumn{5}{c}{1.25$\Msun$, $Z = 0.01$ scaled-solar model.} \\ \tableline
0.26$^{c}$ & 0.0078$^{c}$ & 2.18 & 73.8 & 1225 & 615 \\ 
0.26 & 0.20   & 1.07 & 73.8 & 249  & 1248 \\
0.60 & 0.20   & 2.27 & 157  & 249  & 1248 \\
\tableline 
\multicolumn{5}{c}{1.8$\Msun$, $Z = 0.01$ scaled-solar model.} \\ \tableline
0.24$^{c}$  & 0.0058$^{c}$ & 3.33 & 120 & 338 & 718 \\ 
0.48$^{d}$  & 0.17$^{d}$ & 2.49 & 231 & 869 & 1843 \\
0.24  & 0.10 & 1.76 & 120 & 642 & 1364 \\
0.24  & 0.15 & 1.40 & 120 & 805 & 1706 \\ 
0.60  & 0.20 & 2.78 & 287 & 966 & 2049 \\
\tableline \tableline

\tablenotetext{a}{Intershell abundance (in mass fractions).}
\tablenotetext{b}{All abundance ratios are by number.}
\tablenotetext{c}{Standard intershell composition.}
\tablenotetext{d}{The composition of the post-AGB star PG1159-035.}

\end{tabular}
\end{center}
\end{table}

\clearpage

%% Two options are available to the author for producing tables:  the
%% deluxetable environment provided by the AASTeX package or the LaTeX
%% table environment.  Use of deluxetable is preferred.
%%

%% Three table samples follow, two marked up in the deluxetable environment,
%% one marked up as a LaTeX table.

%% In this first example, note that the \tabletypesize{}
%% command has been used to reduce the font size of the table.
%% We also use the \rotate command to rotate the table to
%% landscape orientation since it is very wide even at the
%% Note also that the \label command needs to be placed

%% This table also includes a table comment indicating that the full
%% version will be available in machine-readable format in the electronic

\bibliographystyle{apj}
% on marmot
%\bibliography{apj-jour,/home/akarakas/biblio/library}
% using the macbook
\bibliography{apj-jour,/Users/amanda/biblio/library}

%% The following command ends your manuscript. LaTeX will ignore any text

\end{document}